\documentstyle[12pt,epsf,a4]{article}
\begin{document}
\enlargethispage{2\baselineskip}
\begin{flushright}
PSI-PR-99-15\\
June 01, 1999
\end{flushright}
\begin{center}
{\large\sffamily\bfseries
STRUCTURE OF THE LIGHT SCALAR MESONS
FROM A COUPLED CHANNEL ANALYSIS OF
THE $S$-WAVE $\pi\pi-K\bar{K}$ SCATTERING
}
\\[5mm]
{\sc 
V.~E.~Markushin 
\footnote{
          Talk given at the Workshop on Hadron Spectroscopy,
          Frascati, March 8-12, 1999
          }
and M.~P.~Locher 
}
\\[5mm]
{\em Paul Scherrer Institute, CH-5232 Villigen, Switzerland}
\\[5mm]
\end{center} 

\baselineskip=14.5pt
\begin{abstract}

The analysis of the poles of the $\pi\pi$ scattering amplitude is 
shown to favour a dynamical origin of the sigma meson: 
it is created by a strongly attractive $\pi\pi$ interaction
in the $J^{PC}=0^{++}I=0$ channel.  In the limit $N_c\to\infty$, the $\sigma$
meson disappears, contrary to the case of conventional $q\bar{q}$ states.
Implications for the search of the scalar $q\bar{q}$ and glueball states 
are discussed. 

\end{abstract}
\baselineskip=17pt  
%

\section{Introduction}
\label{Intro}

   The lightest scalar meson, presently listed by the PDG\cite{RPP98}
under the name $f_0(400-1200)$, known as the sigma meson, is a subject of
long lasting debates. 
On the conservative side, the notion of the $\sigma$ meson simply refers to     
the well established fact that the $\pi\pi$ scattering phase $\delta_0^0(M)$ 
in the $J^{PC}=0^{++}I^G=0^+$ channel strongly rises 
between the $\pi\pi$ threshold and the $f_0(980)$ region 
(see Ref.\cite{Ryb} and references therein).    
   Brave people speak about positions of the $\sigma$ pole(s) of the S-matrix;
however, they have differing opinions about the dynamical origin of this pole.
   In this paper we try to elucidate the nature of the lowest
scalar--isoscalar states. The dynamical origin of the $\sigma$ meson is
discussed in Sec.\ref{SecSigma}, the interplay of different 
scalar--isoscalar states is considered in Sec.\ref{SecCCM}, 
and conclusions are summarized in Sec.\ref{SecConcl}.

\section{Remarks on the model dependence of the $\sigma$-meson results}
\label{SecSigma}

\begin{table}[bht]
\caption{\label{Tabsigma}
The poles of the $S$-matrix in the $\sqrt{s}$-plane (GeV)
corresponding to the $\sigma$ resonance.  
The errors, when available, are shown in parentheses. 
In case when several solutions are found, the average value and 
spread are given.}
\begin{center}
{\small  
\begin{tabular}{lc|c} \hline
 Model &  Ref.         & $M_{\sigma}-i\Gamma_{\sigma}/2\ $ (sheet II)
\\
\hline
 $\rho$ exchange model & \cite{BL71}  & $0.460 - i0.338$
\\
 dispersion relation analysis & \cite{BFP72}  & $0.48(12) - i0.47(15)$
\\
 unitarized meson model & \cite{BRM86}  & $0.47 - i0.21$
\\
 meson exchange model   & \cite{ZB94}  & $0.370 - i0.356$
\\
 linear $\sigma$ model  & \cite{AS94}  & $0.42  - i0.37 $
\\
 coupled channel model  & \cite{KLM94} & $0.506 - i0.247$
\\
 meson exchange model   & \cite{JPHS95}& $0.387 - i0.305$
\\
 unitarized quark model  & \cite{To95,To96}
                                      & $0.47  - i0.25 $
\\
 coupled channel model  & \cite{KLL9798} 
                                      & $0.54(2) - i0.25(2)$ 
\\
 coupled channel model    & \cite{Ish97}  & $0.60(3) - i0.20(3)$
\\
 unitarized chiral perturbation theory (ChPT) 
                          & \cite{OO97}  & $0.470 - i0.179$
\\
 unitarized ChPT  & \cite{OO98}        & $0.442 - i0.227$
\\
 coupled channel model  & \cite{LMZ98} & $0.42 - i0.21$
\\
\hline
 M-matrix (resonance approx.)  & \cite{PAB73}  & $0.66(10) - i0.32(7)$
\\
 K-matrix (resonance approx.)  & \cite{Es79}  & $0.8(1) - i0.5(1)$
\\
 K-matrix (resonance approx.)  & \cite{AMP87}
                                       & $0.87 -i0.37 $
\\
 resonance approx. of the Jost function & \cite{MP93}
                                       & $\sim(1.0 -i0.35)$
\\
 K-matrix (resonance approx.)  & \cite{CBC95a}  & $\sim(1.1 -i0.3)$
\\
 K-matrix (resonance approx.)  & \cite{CBC95}   & $0.4 -i0.5$  
\\
\hline
\end{tabular}
} 
\end{center}
\end{table}

   It is commonly accepted that the rise of the scattering phase 
$\delta_0^0(M)$ in the $\sigma$ region is due to some broad 
resonance.  The problem is whether the position of the corresponding 
pole and its physical origin can be reliably determined. 
   Table 1 lists the properties of the lowest $J^{PC}=0^{++}I^G=0^+$
state found in different models\footnote{%
The models restricted to the region above 1~GeV 
(see \cite{AAS97} and references therein) are not included.}  
of the $\pi\pi$ interaction. 
   Many explicitly dynamical models (listed in the upper part of the table) 
have the "light" $\sigma$ meson in the mass region about 0.5~GeV. 
There is a fair agreement between the pole positions within this "light" $\sigma$ 
group, and the spread of results can be taken as an estimate of the systematic 
errors of the $\sigma$ mass and width.  
   In contrast to these results, the multi--resonance K-matrix fits 
tend to produce higher mass values.   
   In order to understand the difference between these models, 
we briefly summarize their main features in the following table.

\begin{center}
{\small 
\begin{tabular}{lll}
\hline
model         &  strength  & weakness
\\
\hline
K-matrix (res. approx.)
              & convenient fit
              & no left hand cut 
\\
\hline
ChPT          & chiral symmetry, crossing
              & no explicit left hand cut 
\\
\hline
Meson exchange
              & dynamical singularities
              & potential approximation
\\
\hline
\end{tabular}
} 
\end{center}

   We can incorporate the most significant elements of all these models 
in a relatively simple framework of the $\sigma$ meson generation. 
This includes: the S-matrix unitarity, 
the constraints following from chiral symmetry, 
and the explicit manifestation of the closest dynamical singularities.  
To combine all these basic features together we use the N/D method and   
describe the $S$-wave scattering amplitude by 
\begin{eqnarray}
     A(s) & = & (s - s_0) \frac{N(s)}{D(s)} \quad .   \label{EqND}
\end{eqnarray}
Here the factor $(s - s_0)$ is due to chiral symmetry 
($s_0=m_{\pi}^2/2$ is the Adler zero\cite{Ad}), 
the function $N(s)$ is analytical in the physical region and has only 
the left--hand (dynamical) cut, 
and the function $D(s)$ has only the right--hand (kinematical) cut 
as required by unitarity.
Given the discontinuity of the amplitude across the left--hand cut, 
the system of integral equations for the functions $N(s)$ and $D(s)$ 
is obtained in the standard way.
  For the discontinuity of the amplitude on the left hand cut, we take 
the {\it closest} dynamical singularity corresponding to  
the Born approximation for the $\rho$ meson $t$-- and
$u$--channel exchanges\footnote{The left hand cut runs from 
$s_L=4m_{\pi}^2-m_{\rho}^2$ to $s=-\infty$. 
The discontinuity function is cut off for $s<s_c$ 
to avoid otherwise needed subtractions.  
The cutoff position $s_c$ is determined from the requirement 
that the $\pi\pi$ scattering length $a_0^0=0.2 m_{\pi}^{-1}$ corresponds to
the physical value of the $\rho\pi\pi$ coupling constant
$g_{\rho\pi\pi}=6.0$ (the value of $s_c$ turns out to be close to
the dynamical threshold at $s=4m_{\pi}^2-4m_{\rho}^2$ corresponding to 
the second order $\rho$--exchange term).}:   
\begin{eqnarray} 
   {\mathrm disc}\ A(s) & = & 
   {\mathrm disc}\ 
   \raisebox{-5mm}{\mbox{\epsfysize=10mm\epsffile{texch1.epsf}}} 
   \ = \  
   g_{\rho\pi\pi}^2 \frac{(2s -4m_{\pi}^2 + m_{\rho}^2)}{8(s-4m_{\pi}^2)} 
   \theta(4m_{\pi}^2-s-m_{\rho}^2) 
   \quad . 
\end{eqnarray}

\begin{figure}[hbt]
\mbox{\hspace*{35mm}(a) \hspace*{45mm}(b) \hspace*{45mm}(c)}\\[-8mm]
\mbox{\hspace*{-10mm}
\mbox{\epsfysize=60mm \epsffile{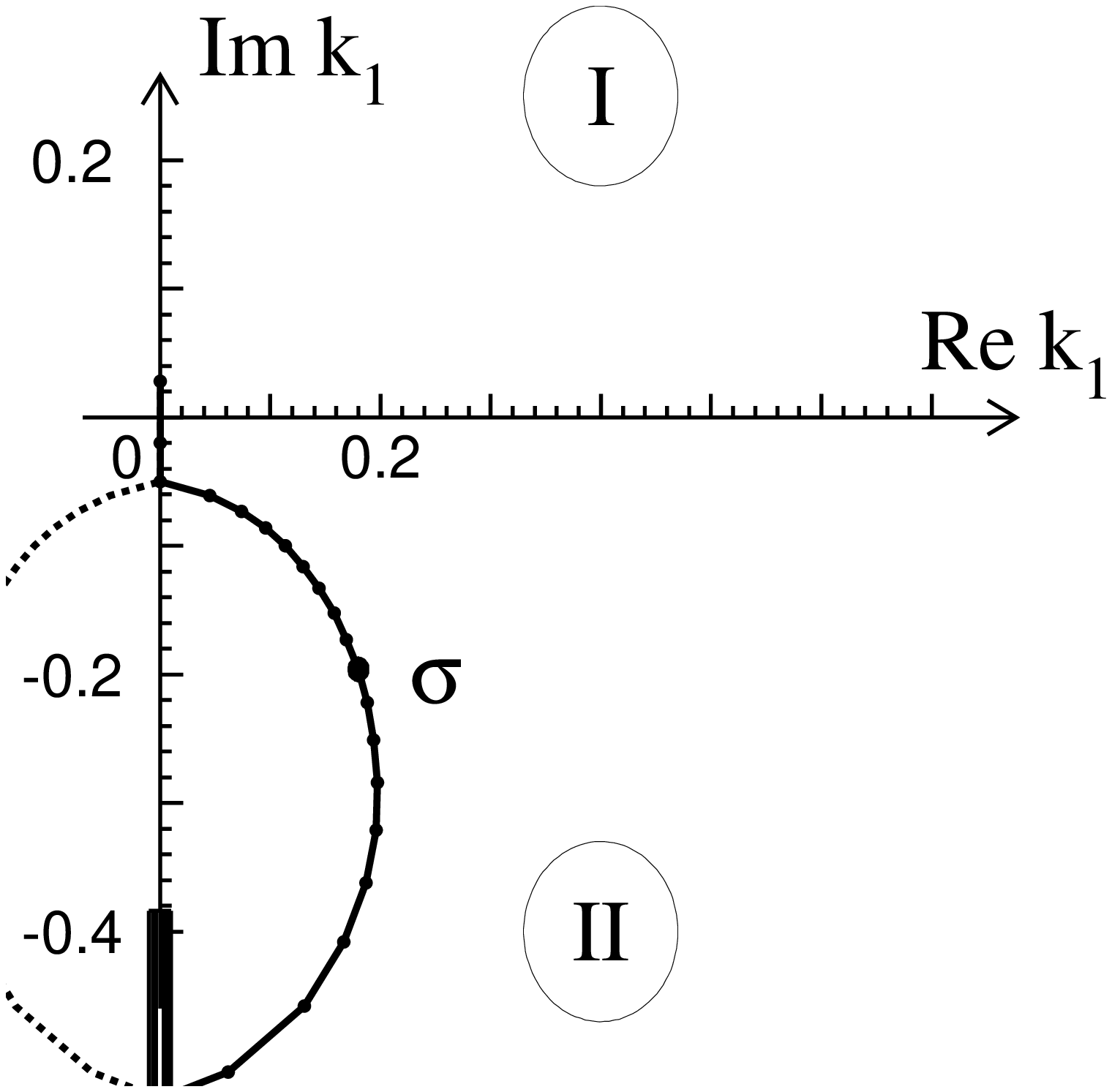}} \hspace*{-10mm}
\mbox{\epsfysize=60mm \epsffile{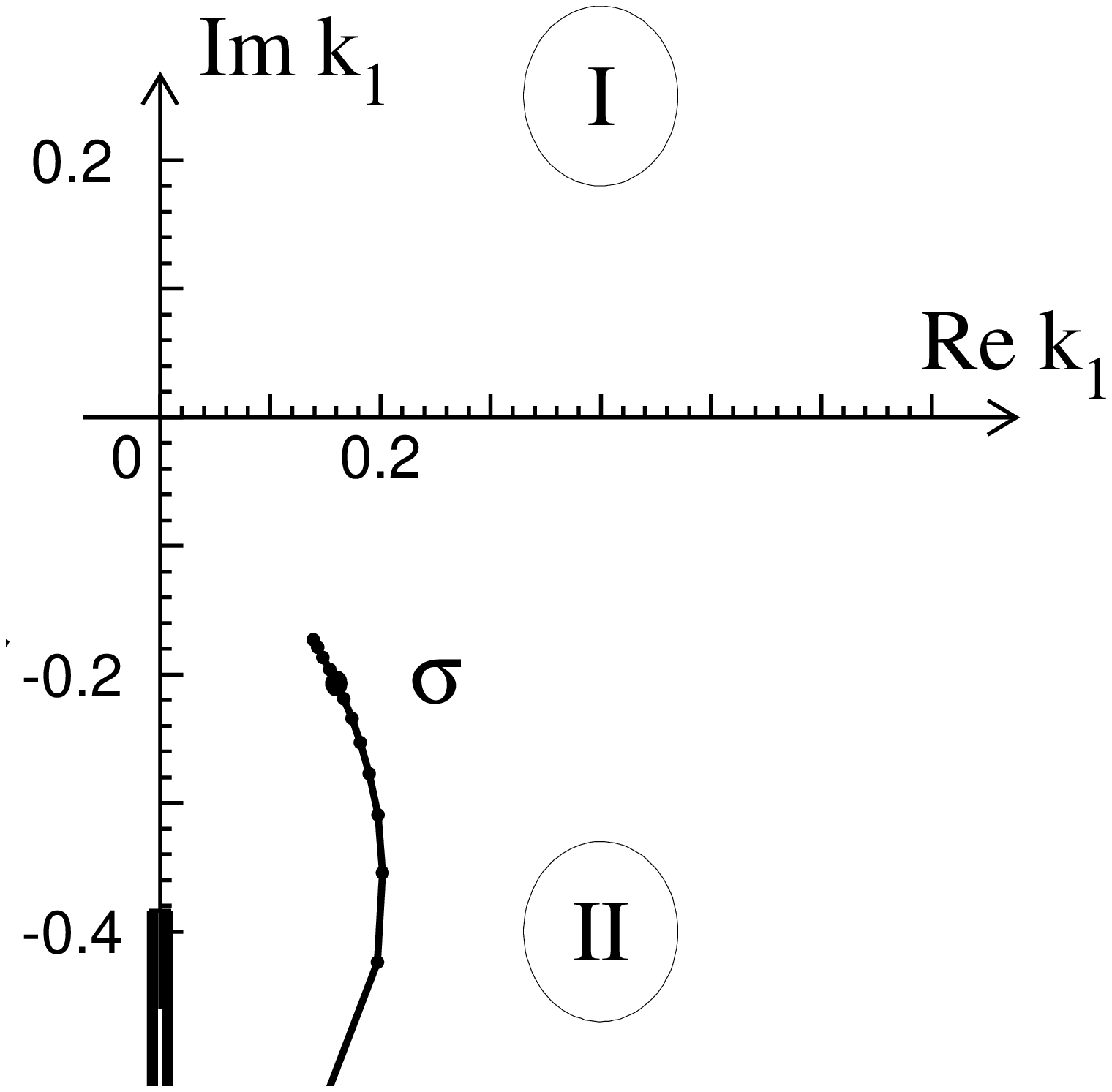}} \hspace*{-10mm}
\mbox{\epsfysize=60mm \epsffile{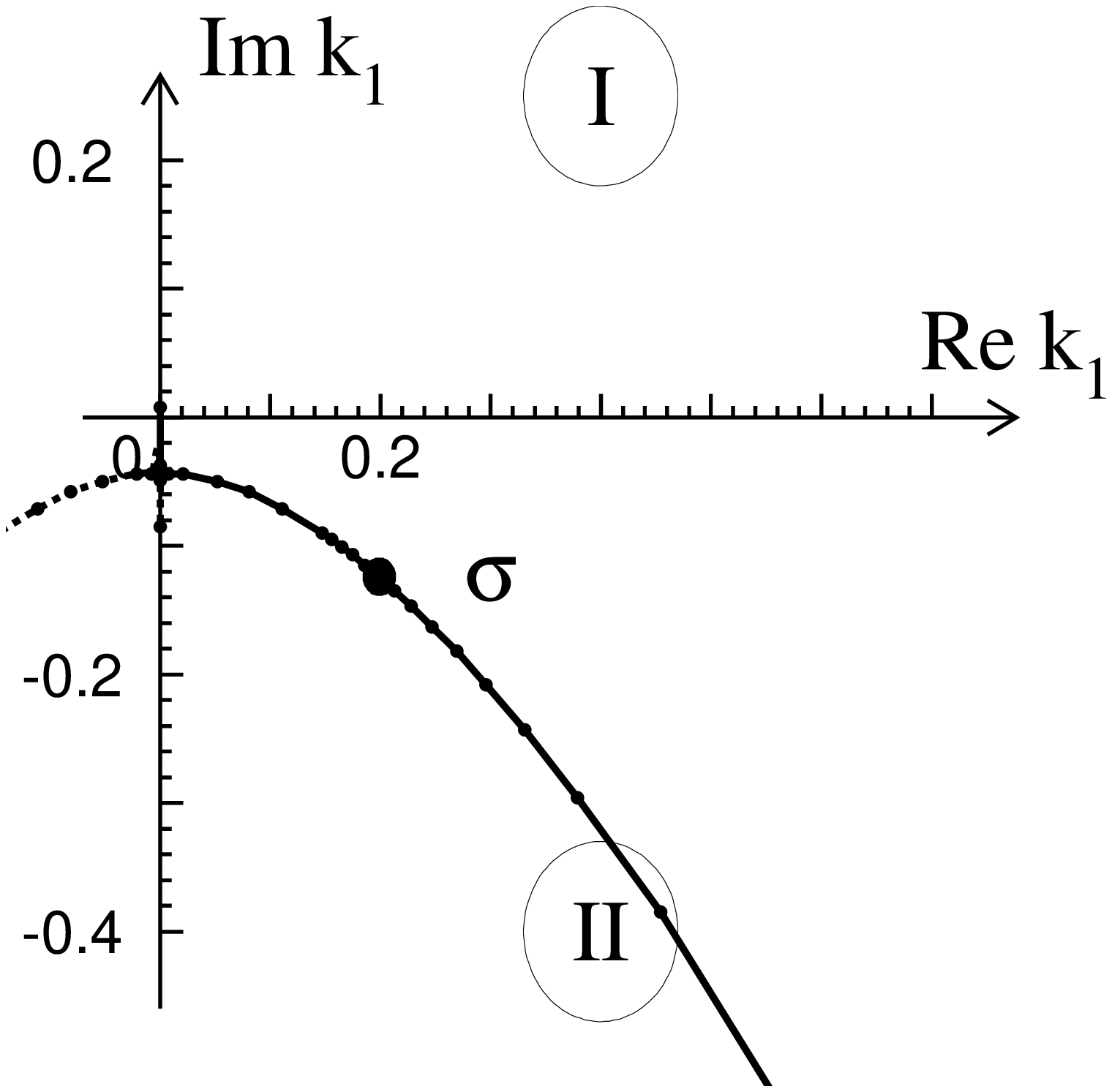}}
}\\[-10mm] 
\caption{\label{Figsigma}
The trajectories of the $\sigma$ meson pole in the complex plane
of the relative $\pi\pi$ momentum (GeV):
(a) the N/D method with the Adler zero and the $\rho$-exchange 
left hand cut, 
(b) the K-matrix unitarization of the $\rho$-exchange Born approximation,
(c) the one loop ChPT unitarized. 
The labels I and II indicate the corresponding Riemann sheets of 
the amplitude as a function of the total energy.   
The physical values are marked with $\protect\bullet$.
}
\end{figure}

   Figure \ref{Figsigma}a shows the trajectories of the
$\sigma$ pole for variable strength of the $\pi\pi$ interaction. 
   As the $g_{\rho\pi\pi}$ coupling constant increases from zero to the
physical value, a pair of poles corresponding to the $\sigma$ meson
emerges from the dynamical cut and moves  
on the second sheet of the complex energy plane. 
For the scattering length $a_0^0=0.2 m_{\pi}^{-1}$
the pole position in the energy plane is
$(M_{\sigma}-i\Gamma_{\sigma}/2)=\mbox{$(0.40-i0.29)$}\;$GeV, in a good agreement
with the more elaborated models listed in~Tab.~\ref{Tabsigma}.
With further increase of $g_{\rho\pi\pi}$, the poles collide at the
Adler zero and then move along the imaginary $k$-axis\footnote{This kind of
resonance trajectories resembles the well known case of potential
scattering in the $P$-wave where the poles collide at threshold,
with the threshold position replacing $s_0$ in (\ref{EqND}).}. 

   From this example, we can draw a more general conclusion by using  
the fact that the limit $g_{\rho\pi\pi}\to 0$ corresponds to the QCD limit of 
large number of colors, $N_c\to\infty$.      
While the masses of the $q\bar{q}$ mesons remain finite and their widths 
become infinitely small at $N_c\to\infty$, the $\sigma$ meson completely vanishes 
in this limit.  In other words, the $\sigma$ meson 
corresponds to a dynamical pole of the S-matrix and, thus, it is fundamentally  
different from the conventional $q\bar{q}$ states (like $\pi$ and $\rho$)  
which correspond to the CDD poles\cite{CDD}.       
  
   The relation of the $\sigma$ meson to the $\pi\pi$ attractive interaction 
generated by the $\rho$--exchange, has been discussed in the literature 
\cite{ZB94,JPHS95}. 
Using the K-matrix unitarization of the Born approximation, as suggested in 
\cite{ZB94}, we obtain the $\sigma$ meson trajectory shown in Fig.\ref{Figsigma}b.  
This simple approximation\footnote{%
Because this approximation breaks down at large $g_{\rho\pi\pi}$, 
the trajectories do not return to the imaginary $k$ axis.} 
produces fairly reasonable results up to the physical value of $g_{\rho\pi\pi}$ and 
clearly demonstrates the importance of the left hand cut in the dynamical generation 
of the S-matrix poles.   

   Turning to the chiral perturbation theory and using  
the one--loop single channel approximation unitarized as in Ref.\cite{OO97}, 
we obtain the $\sigma$ trajectory shown in Fig.\ref{Figsigma}c. 
   For increasing strength of the $\pi\pi$ interaction, the pole
trajectories  behave similar to the case in Fig.\ref{Figsigma}a 
because both models satisfy the same constraints imposed by the chiral symmetry 
(they have the same Adler zero). 
However, the ChPT trajectories go to infinity when the interaction strength 
goes to zero (the pion decay constant $f\to\infty$). 
  Comparing the different cases shown in Fig.\ref{Figsigma} one can conclude 
that the role of the left hand cuts in the ChPT framework must be further 
investigated.  It is known\cite{Wein,BKM91} that the $\rho$-meson exchange term 
receives a strong cancellation in the leading order of the ChPT expansion. 
On the other hand, the discontinuity of the $\rho$ exchange amplitude  
across the left hand cut is sufficient for generating poles in the complex 
plane.  While the Pad\'e approximation of the ChPT series can easily produce 
the $\sigma$ pole, it is more difficult to reconstruct the left hand cut 
in this approach.  The simplified version of the ChPT in Fig.\ref{Figsigma}c 
just puts all other singularities (including the left hand cut) to infinity. 
It is rather plausible that, if proper dynamical cuts are taken into account,  
the $\sigma$ trajectories will originate from the left hand cut rather 
than from infinity. 
        
   A similar situation seems to occur in the multichannel K-matrix fits 
based on multiresonance approximation (Table \ref{Tabsigma}) where no dynamical 
cuts are taken into consideration and the total number of S-matrix poles 
does not depend on the coupling constants.  While these models can provide 
good fits\cite{AAS97} of the experimental data in the physical region, the lack 
of theoretical information related to the left hand cut apparently distorts 
the nature of the $\sigma$ origin.  
In this situation, the rising scattering phase in the 
$\sigma$ region is built up by heavier resonances.

\section{The $\pi\pi - K\bar{K}$ Coupled Channel Model}
\label{SecCCM}

  Another important aspect of the $\sigma$ meson dynamics is its interplay 
with further $J^{PC}I^G=0^{++}0^+$ states involving  
the $f_0(980)$, the $q\bar{q}$ states, and the scalar glueball.    
Here we exploit an exactly solvable coupled channel model\cite{LMZ98} to 
address some of these questions. 
  The following interactions are included in this model: a separable diagonal
$K\bar{K}$ potential producing a weakly bound state\footnote{The existence  
of such a state in a single channel approximation was suggested in different 
models, see e.g. \cite{WI90}.}, 
a separable transition potential $V_{\pi\pi-K\bar{K}}$ 
(representing $K^*$-exchange, e.g.) which
couples the $\pi\pi$ and $K\bar{K}$ channels, and broad resonances in the
$q\bar{q}$ channel which represent background and are coupled to
the $\pi\pi$ and $K\bar{K}$ channels. The chiral symmetry constraints 
are imposed on the model parameters, so that the Adler zero is always located 
at the right position.  
The model parameters have been determined from the fit of the $\pi\pi$ 
and $K\bar{K}$ scattering data, see Ref.\cite{LMZ98} for details. 
  This model allows one to study the appearance of dynamical coupled channel
poles in the case of physical parameters, where the number of poles close to the
physical region exceeds the number of the bare states.   The
interpretation of the singularities, including the important role of the 
dynamical ones, is elucidated by studying the trajectories of the
$S$-matrix poles as a function of the strength of the channel couplings.

\begin{figure}[htb]
\mbox{\hspace{50mm}(a)\hspace{65mm}(b)}\\[-8mm]
\mbox{
\hspace{-15mm}
\mbox{\epsfysize=8cm \epsffile{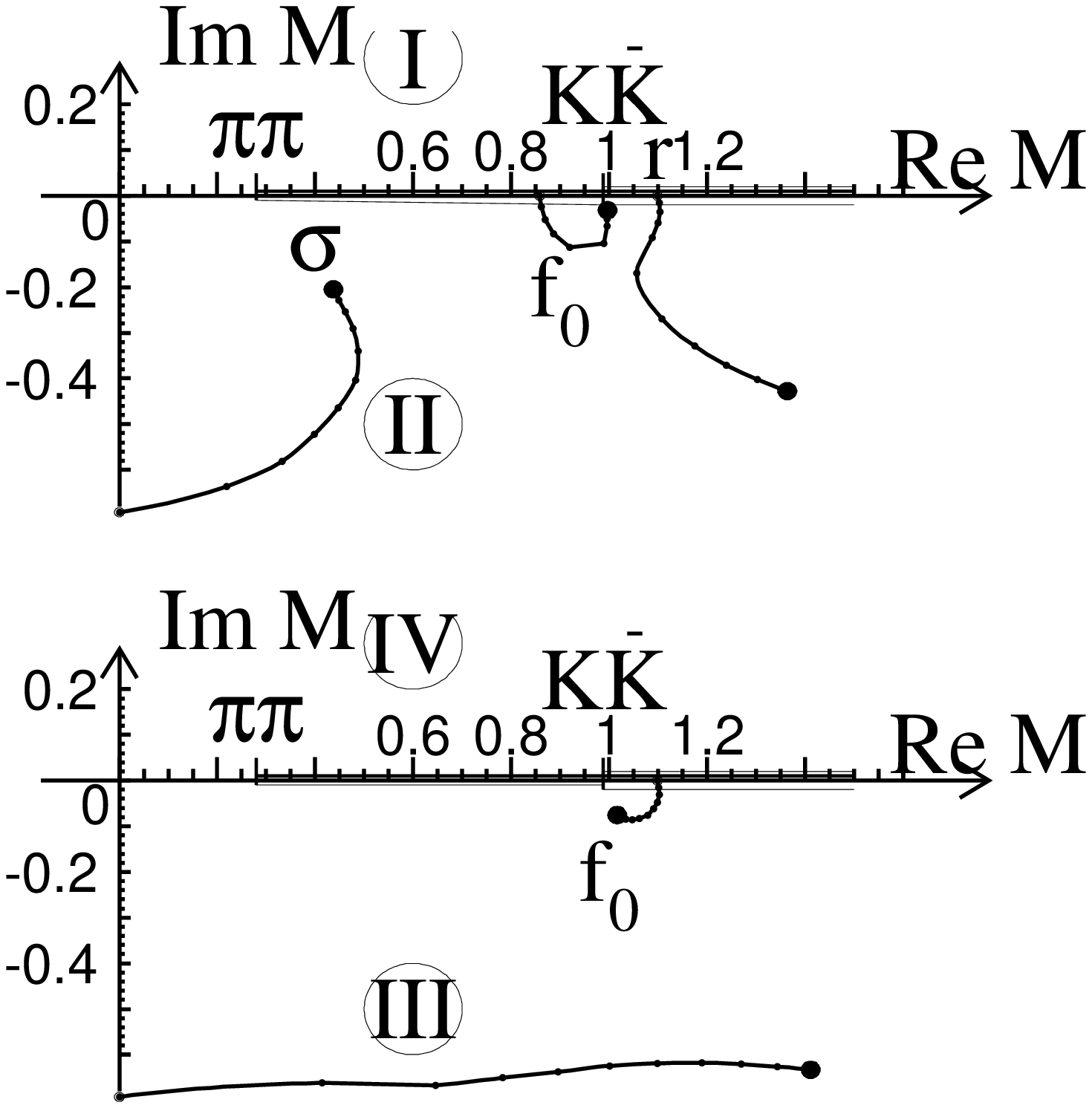}}
\hspace{-10mm}
\mbox{\epsfysize=8cm \epsffile{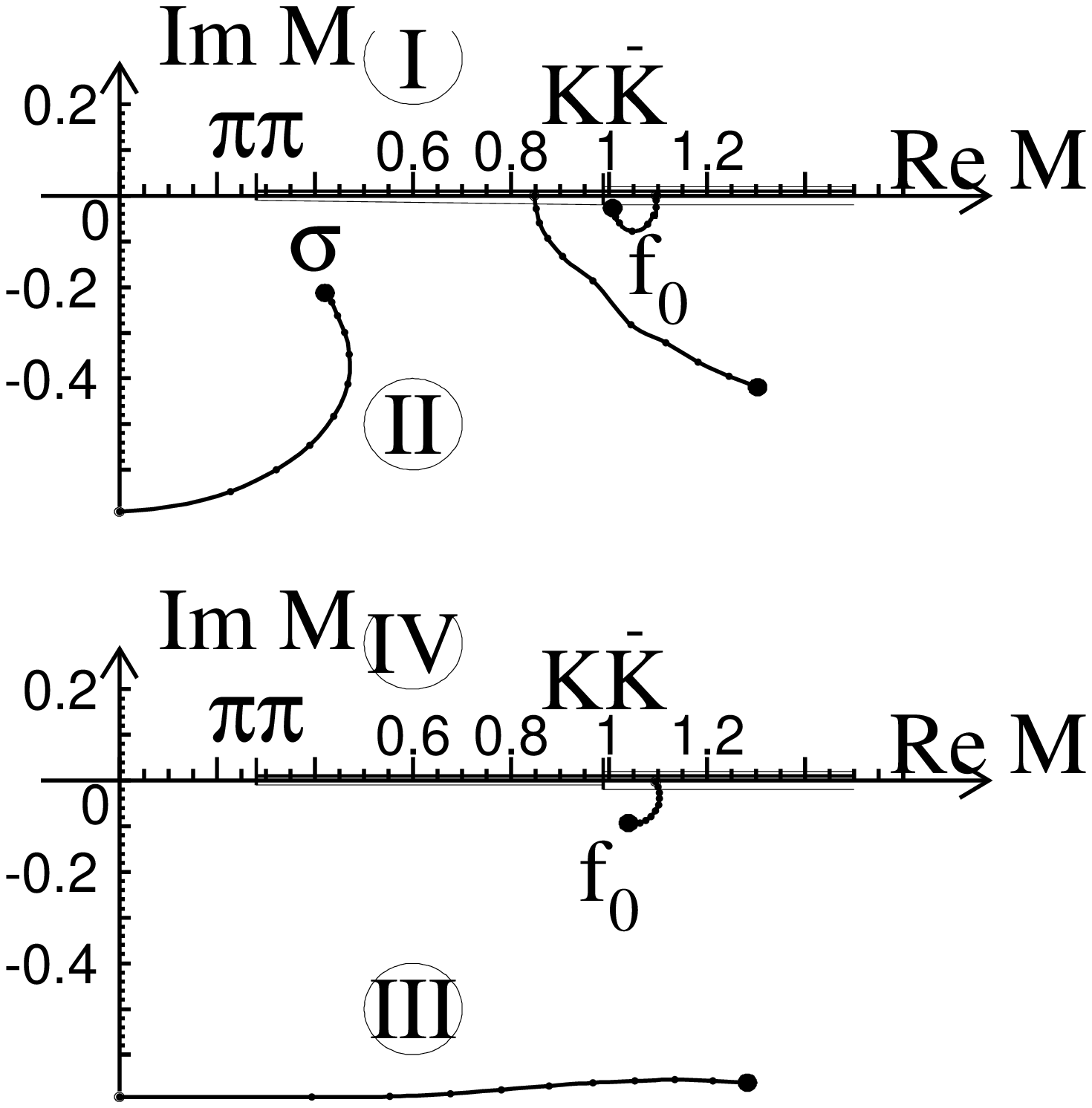}}
}\\[-15mm] 
\caption{\label{Figtrajs}
The trajectories of the S-matrix poles in the $M=\sqrt{s}$ plane 
for the coupled channel model\protect\cite{LMZ98}. 
The pole positions for the physical values are marked with 
$\protect\bullet$. 
The $\sigma$ pole on the sheet II originates from the dynamical 
singularity. 
The $f_0(980)$ pole on the sheet II originates from the $K\bar{K}$ bound 
state in case (a) and from the $q\bar{q}$ bare state in case (b).  
}
\end{figure}

  Figure \ref{Figtrajs} shows two typical cases of the trajectories of 
the S-matrix poles in the complex plane $M=\sqrt{s}$, the couplings between 
the channels having been varied between zero and their physical values.   
  In both cases, the $\sigma$ meson is generated dynamically on the second 
sheet of the complex energy plane by the strong coupling
$q\bar{q} \leftrightarrow \pi\pi$ resulting in an effective attractive
interaction in the $\pi\pi$ channel below the $q\bar{q}$ state. 
The singularity of the form factor in the $(q\bar{q})\pi\pi$ vertex 
plays the same role as the left hand cut in the discussion in Sec.2.  
Thus the situation in the vicinity of the $\pi\pi$ threshold is similar 
to that of the single channel approximation discussed above.  

  The situation for the $f_0(980)$ near the $K\bar{K}$ threshold is much richer: 
there are two different possibilities shown in Figs.\ref{Figtrajs}a,b. 
In case (a), the physical pole on the sheet II corresponding to the $f_0(980)$ 
originates directly from the weakly bound $K\bar{K}$ state, 
and the $(q\bar{q})$ state on the sheet II develops a large imaginary part. 
In case (b), the pole originating from the $(q\bar{q})$ state is attracted 
to the $K\bar{K}$ threshold and corresponds to the narrow $f_0(980)$ state. 
It can be demonstrated using the probability sum rules\cite{LMZ98} that in the 
latter case we have rearrangement of the spectrum where the initial $q\bar{q}$ 
state turns into a molecular $K\bar{K}$ state with a small admixture of 
the original $q\bar{q}$ component.    
 
  Figure \ref{Figtrajs} also demonstrates that in both cases there are two poles 
close to $K\bar{K}$ threshold (on the sheets II and III) in agreement with many fits 
to the data.  This result does not contradict the $K\bar{K}$ molecular origin 
of the $f_0(980)$  
because the conventional wisdom of having only one pole near threshold 
for a molecular state is based on perturbation theory and does not apply 
to the strong coupling case. 
The additional pole originates from the $q\bar{q}$ bare state and moves close to 
the $K\bar{K}$ threshold due to the strong attraction in the $K\bar{K}$ channel 
while the $q\bar{q}$ component dissolves in the continuum producing a broad 
resonance.

\section{Conclusion}
\label{SecConcl}

   Our results lead to the following picture of the lowest scalar--isoscalar 
mesons. 
The lightest state is the broad $\sigma$ meson which corresponds to the 
S-matrix pole at $(M_{\sigma} - i \Gamma_{\sigma} \approx (0.4 - i 0.2)\;$GeV. 
This pole has a dynamical origin: it is produced by the attractive interaction 
in the $\pi\pi$ system.  The $\sigma$ meson is fundamentally different from 
the $q\bar{q}$ resonances, in particular, it vanishes in the limit $N_c\to\infty$  
where the $q\bar{q}$ resonances turn into narrow states. 
 
   The second state is the $f_0(980)$ near the $K\bar{K}$ threshold. It corresponds 
to two S-matrix poles on the sheets II and III in the two-channel case (there 
will be more poles if extra channels are added to the $\pi\pi$ and $K\bar{K}$ 
channels). 
This state is also of dynamical nature because it originates from a $K\bar{K}$ 
molecular state embedded in the $\pi\pi$ continuum.  This state vanishes 
in the limit $N_c\to\infty$ when the $K\bar{K}$ interaction becomes weak. 

Therefore, the $q\bar{q}(0^+)$ states are expected in the mass region  
$M>1\;$GeV where several $J^{PC}=0^{++}I=0$ resonances have been 
found\cite{RPP98}.  
It is rather gratifying to have the two lowest states accommodated beyond 
the $q\bar{q}$ family because otherwise there would appear to be too many 
candidates to be identified with the $(u\bar{u}+d\bar{d})$, $s\bar{s}$, 
and the scalar glueball states (or the corresponding mixtures of them).     

\section*{Acknowledgments}

The participant of WHS-99 (V.M.) thanks all the organizers, in particular 
T.~Bressani, for the interesting and stimulating workshop.  
Fruitful discussions with R.~Kami\'{n}ski, L.~Montanet, and J.A.~Oller 
are very much appreciated.


%
\end{document}